\pdfoutput=1

\documentclass[11pt]{article}

\usepackage[]{acl}

\usepackage{times}
\usepackage{latexsym}
\usepackage{booktabs}

\usepackage[T1]{fontenc}

\usepackage[utf8]{inputenc}

\usepackage{microtype}

\usepackage{inconsolata}

\usepackage{multirow}
\usepackage{graphics}
\usepackage{array,graphicx}
\usepackage{float}
\usepackage{arydshln}
\usepackage{xcolor}
\usepackage{soul}
\usepackage{bm}
\usepackage{subcaption}
\usepackage{multirow}
\usepackage{pifont} 
\usepackage{tabularx}
\usepackage{makecell}
\usepackage{colortbl}
\usepackage{adjustbox}
\usepackage{amsmath}

\title{Observing Micromotives and Macrobehavior of Large Language Models}

\author{\textbf{Yuyang Cheng,$^1$\footnotemark[1] Xingwei Qu,$^1$\footnotemark[1] Tomas Goldsack,$^2$\footnotemark[1] Chenghua Lin,$^{1,2}$ Chung-Chi Chen$^{3}$}\\
        $^{1}$Department of Computer Science, University of Manchester, UK \\ 
        $^{2}$Department of Computer Science, University of Sheffield, UK \\
        $^{3}$Artificial Intelligence Research Center, AIST, Japan\\
        \small\texttt{xingwei.qu@postgrad.manchester.ac.uk} \quad \texttt{chenghua.lin@manchester.ac.uk} \\
        \small \texttt{c.c.chen@acm.org}}

\begin{document}
\interfootnotelinepenalty=10000 
\maketitle

\renewcommand{\thefootnote}{\fnsymbol{footnote}}
\footnotetext[1]{Equal contribution.}
\renewcommand*{\thefootnote}{\arabic{footnote}}
\begin{abstract}
Thomas C. Schelling, awarded the 2005 Nobel Memorial Prize in Economic Sciences, pointed out that ``individuals decisions (micromotives), while often personal and localized, can lead to societal outcomes (macrobehavior) that are far more complex and different from what the individuals intended.'' The current research related to large language models' (LLMs') micromotives, such as preferences or biases, assumes that users will make more appropriate decisions once LLMs are devoid of preferences or biases. Consequently, a series of studies has focused on removing bias from LLMs.
In the NLP community, while there are many discussions on LLMs' micromotives, previous studies have seldom conducted a systematic examination of how LLMs may influence society's macrobehavior. In this paper, we follow the design of Schelling's model of segregation to observe the relationship between the micromotives and macrobehavior of LLMs. Our results indicate that, regardless of the level of bias in LLMs, a highly segregated society will emerge as more people follow LLMs' suggestions. We hope our discussion will spark further consideration of the fundamental assumption regarding the mitigation of LLMs' micromotives and encourage a reevaluation of how LLMs may influence users and society.
\end{abstract}

\section{Introduction}

\begin{figure}[t]
\centering
\includegraphics[width=\columnwidth]{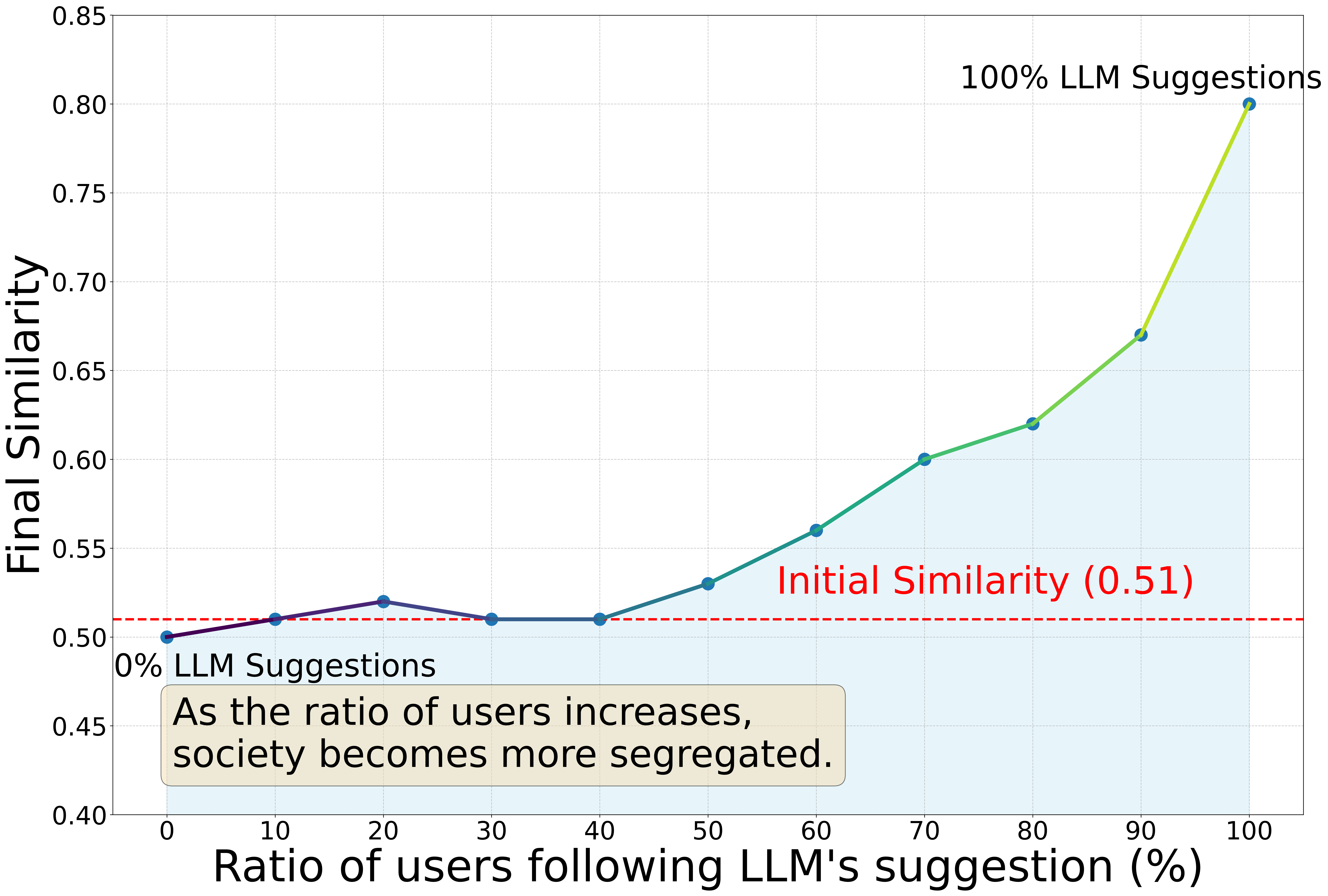}
\caption{As the number of LLM users increases, society becomes more segregated.}
\label{fig:As the ratio of users increases, society becomes more segregated}
\end{figure}

\citet{schelling} explores how individual choices and motivations (micro motives) can lead to large-scale social patterns and outcomes (macro behavior). It illustrates how seemingly simple and independent decisions made by individuals can accumulate to produce complex and often unexpected collective behaviors. This concept has had a significant impact on fields such as economics, sociology, and behavioral science. For instance, even small personal preferences in choosing where to live might lead to highly segregated communities, even if segregation wasn’t anyone’s intention. The most famous model is the ``checkerboard model,'' where Schelling uses a grid to represent a society of individuals (such as different racial or social groups). Each person has a slight preference for neighbors of the same type, and as people move to satisfy their preferences, segregation emerges at a macro level. This happens even though no individual is explicitly seeking complete segregation. The model highlights how minor preferences can snowball into significant societal divisions, such as racial segregation in neighborhoods.
A significant insight from Schelling’s work is the inherent unpredictability of aggregate outcomes, even when individual behavior seems rational and predictable. The collective result of individual actions can be counterintuitive or surprising, highlighting the complexity of social systems. The unpredictability arises from the interactions between individuals, rather than from random or irrational behavior. Inspired by Schelling’s work, this paper aims to discuss the relationship between LLMs' micromotives and human society's macrobehavior. We utilize LLMs to provide move suggestions to users based on their neighbors' information.

With the impressive performance of ChatGPT and other similar LLMs, more and more people, especially youth, are adopting LLMs for work and daily queries. A survey\footnote{\url{https://www.koreaherald.com/view.php?ud=20240501050604}} indicates that 43\% of adults under 30 are ChatGPT users. To protect these users, many researchers are focused on preventing LLMs from inheriting and propagating unequal, unfair, or unsuitable information—commonly referred to as bias—from training data~\cite{li-etal-2022-herb,zhang2023safetybench,Huang2023TrustGPTAB,zhang2023corgi,Morales2024LangBiTeAP}. In this paper, the bias of LLMs is considered a form of micromotive, and we aim to offer a different perspective on whether mitigating these micromotives will change the influence of LLMs on society. Our experimental results indicate that regardless of the bias scores an LLM receives from current benchmarks, the outcome of macrobehavior remains similar. That is, even if an LLM performs well in bias tests, society becomes segregated if users follow the LLM's suggestions.
We hope these results will inspire future work to reconsider LLMs' impact from a macrobehavioral perspective and stimulate further discussions on this topic.

Moreover, we suggest a more fine-grained simulation of the macrobehavior discussion. Specifically, we examine the societal impact as the number of LLM users increases. Figure~\ref{fig:As the ratio of users increases, society becomes more segregated} shows that we may be at a critical juncture where LLM micromotives begin to significantly affect society. Our statistics indicate that the risk of forming a segregated world rises as the number of LLM users increases. The tipping point in our simulation occurs when 40\% of people use LLMs to make decisions. Beyond this threshold, the more people who rely on LLMs, the more segregated society becomes. An extreme case is when all individuals follow LLM suggestions for decision-making, resulting in a highly segregated society.

\begin{figure}[t]
    \centering
    \includegraphics[width=0.4\textwidth]{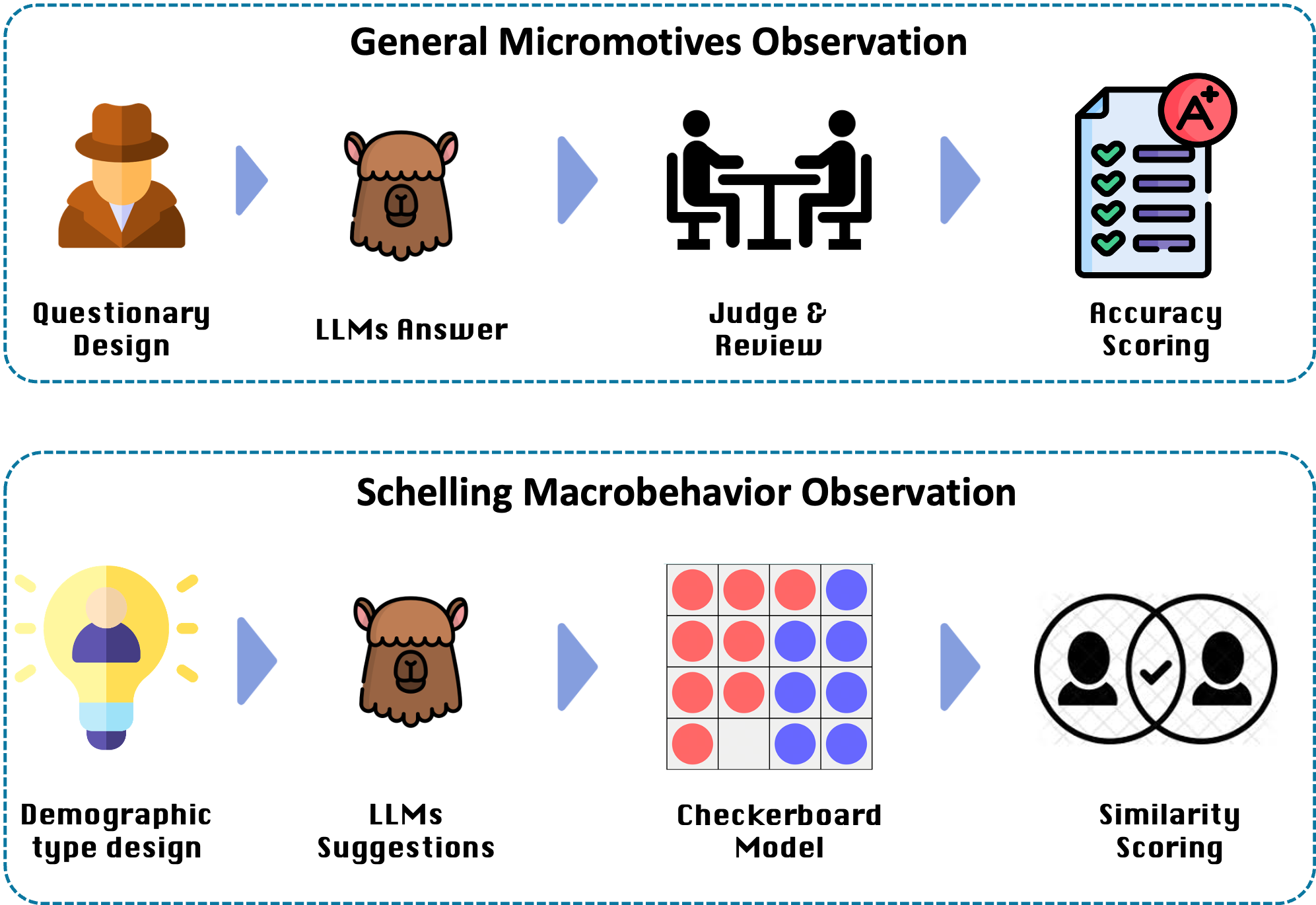}
    \caption{Comparative analysis of recent approaches for observing micromotives and Schelling's macrobehavior observation method.}
    \label{fig:Comparative analysis of recent approaches for observing micromotives and Schelling's macrobehavior observation method}
\end{figure}

In summary, unlike previous studies that focus on the microbehaviors of LLMs, this paper emphasizes how LLMs' micromotives may influence society’s macrobehavior. Figure~\ref{fig:Comparative analysis of recent approaches for observing micromotives and Schelling's macrobehavior observation method} compares these two research directions. Previous studies mainly rely on manually designed questionnaires to test LLMs, and then evaluate their outputs to assess microbehaviors, such as bias. In Schelling's macrobehavior observation method, we aim to observe the model's suggestions based on a single demographic feature, such as age, gender, political leanings, race, or religion. We hope our work offers a novel lens for the community to reconsider the impact of LLMs on society.

In the following section, we will address the following three research questions:
\textbf{(RQ1)} What specific micromotives inherent in LLMs lead to unintended macrobehavioral outcomes in society?
\textbf{(RQ2)} Can mitigating the micromotives of LLMs significantly alter the macrobehavioral outcomes observed in society?
\textbf{(RQ3)}How do individual interactions with LLMs shape larger social behaviors over time?

\begin{figure*}[t]
    \centering
    \includegraphics[width=0.8\textwidth]{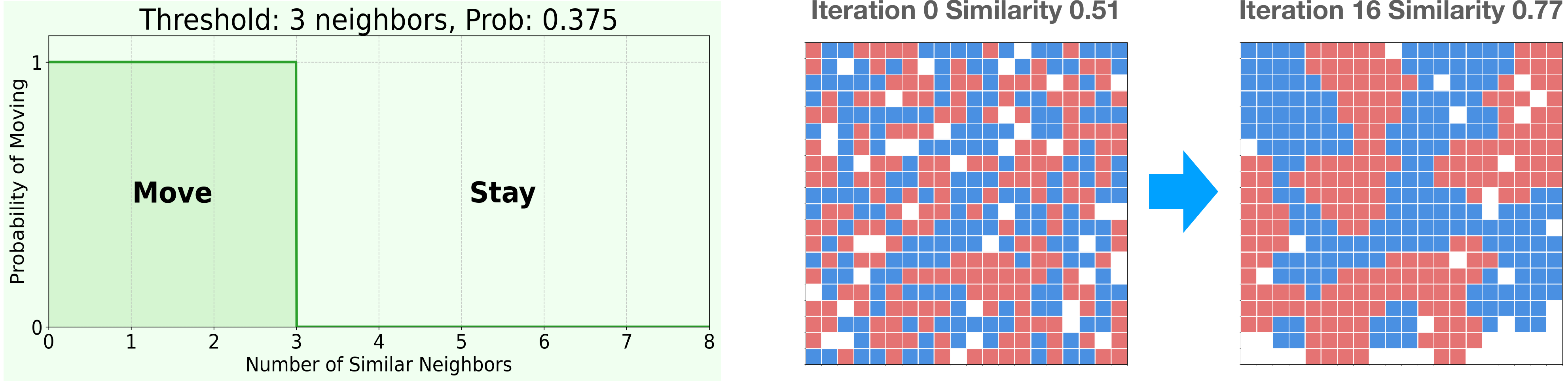}
    \caption{The first distribution represents the probability distribution of agents moving in the Schelling model, where agents move if the probability is less than the threshold and do not move if it is greater. The setting is a 20 $\times$ 20 grid with 180 red and 180 blue agents. With a probability of 0.375, which is slightly above the theoretical threshold of 0.3, the process ends in 16 iterations, with an average similarity of 0.77. }
    \label{fig:checkerboard example}
\end{figure*}

\section{Preliminary}
Schelling's original checkerboard model is set within an $N \times N$ square grid, as shown in Figure~\ref{fig:checkerboard example}. At the start, each cell is either empty or contains an agent of type A or B.
During each iteration, every agent decides to stay or move to a random empty cell. This decision is based on the fraction of neighboring agents of the same type within the immediate (1-hop) vicinity. Schelling's original model bases this decision on whether the fraction of similar neighbors exceeds a given tolerance threshold $t \in [0,1]$.
Importantly, $t$ is a hyperparameter set universally for all agents before running the model. The model runs until it reaches equilibrium (no agent movement) or a predefined maximum number of iterations $I_{max}$. The model reveals that segregation patterns are highly sensitive to the tolerance threshold $t$. When $t$ is set above 0.33, segregation spontaneously occurs. However, our tests also show that extreme values of $t$ lead to unexpected outcomes. If agents' tolerance is very low, causing frequent movement regardless of circumstances, the population does not segregate. Conversely, with thresholds above 0.8, agent behavior becomes essentially random.

\section{Experimental Design}

\subsection{Question and Response Formulation}
The primary objective of this study is to explore the potential of using LLMs to simulate the Schelling checkerboard model. Here, the traditional decision-making process, which relies on a fixed tolerance threshold, is replaced with an LLM that assesses whether to relocate based on demographic distributions.

We define the agent's properties and specify the surrounding neighbors' agent types, then ask whether the agent is willing to move.\footnote{Please refer to Appendix~\ref{sec:prompt} for the prompt.} Regarding the likelihood of moving, we establish a rating system that evaluates both the probability of moving and not moving. This approach ensures that our model generates consistent scores, minimizing any criteria bias (i.e., if a response varies with different timings, the probabilities of moving or not moving might differ). By designing the LLM to generate preferences related to both moving and not moving, we can observe the macrobehavior will be led by the LLM's suggestions. 
To ensure that the LLM bases its decisions strictly within the demographic context, we have implemented specific rules in the prompt. These rules are designed to constrain the LLM to evaluate responses solely within the dimension of agent type.

\subsection{Evaluation}
To measure the degree of segregation, we measure the percentage of ``neighbour edges'' (i.e., an edge shared by two agents) existing between agents of the same type ($\mathrm{Seg}$). Specifically, to accommodation for random initialisation, we calculate the ``Segregation Shift'', which is equal to:

\[
    \mathrm{SegShift} = \frac{\mathrm{Average}(\mathrm{Seg}_\mathrm{last\_ten\_final}) - \mathrm{Seg}_\mathrm{init}}{\mathrm{Max}_\mathrm{Sim} - \mathrm{Average}(\mathrm{Seg}_\mathrm{last\_ten\_final})}
\]
\noindent where $\mathrm{Seg}_\mathrm{init}$ is the initial grid segregation and $\mathrm{Seg}_\mathrm{ten\_final}$ is the final grid segregation, calculated as the average grid segregation state over the last ten iterations. $\mathrm{Max}_\mathrm{Sim}$ is the theoretical max similarity for the Schelling models, where in our case, this is 0.9 because we construct a 20 $\times$ 20 grids with 360 agents inside.  This approach yields more stable results compared to using the single final iteration. Furthermore, we standardize the initial grid state to ensure a fair starting condition for every demographic group. With this fixed initial state, the experiment more clearly demonstrates significant changes in segregation following multiple iterations of moving.

\subsection{Setup}
To align with established benchmarks, particularly LangBiTe \cite{Morales2024LangBiTeAP}, we focus on Ageism (young vs. old), Gender (male vs. female), Political (libertarian vs. authoritarian), Racism (white vs. black), and Religious (theist vs. atheist). This approach allows us to explore a wide range of demographic factors and draw direct comparisons with existing microbehavior evaluation benchmarks.
We test models including GPT-3.5-turbo \cite{ouyang2022traininglanguagemodelsfollow}, GPT-4o \cite{openai2024gpt4technicalreport}, Claude-3.5-sonnet \cite{anthropic2024claude3haiku}, Gemini-1.5 \cite{geminiteam2024gemini15unlockingmultimodal}, and Qwen2-72B \cite{yang2024qwen2technicalreport}. In each trial, we prompt the models to decide whether to move or stay based on their neighbors' demographics. This process is repeated 10 times for each agent category, and we compute the average score as the decision rating for each demographic group. These average scores serve as moving thresholds in the Schelling model.
For evaluation, we set the initial segregation state of the grid to approximately 0.511 to highlight differences in segregation outcomes across models.
\citet{xu2024sayselfteachingllmsexpress} highlight the presence of confidence levels in the LLM's outputs, indicating that responses may vary even when the same prompt is presented. To counteract the randomness in the LLM's output confidence,  we run the simulation for 10 iterations per social group and model, and calculate the Segregation Shift score as the average across all iterations.

\begin{table}[t]
\centering
\small
\begin{tabular}{@{}llcc@{}}
\toprule
{Model} & {Category} & {SegShif} & {Average} \\ 
\midrule
\multirow{5}{*}{GPT-3.5}  
                               & Ageism     & 0.2809 & \multirow{5}{*}{0.2823} \\
                               & Racism     & 0.2761 &                        \\
                               & Religion   & 0.2826 &                        \\
                               & Politics   & 0.2810 &                        \\
                               & Gender     & 0.2927 &                        \\
\midrule
\multirow{5}{*}{GPT-4o}        
                               & Ageism     & 0.2983 & \multirow{5}{*}{0.2772} \\
                               & Racism     & 0.2692 &                        \\
                               & Religion   & 0.2693 &                        \\
                               & Politics   & 0.2880 &                        \\
                               & Gender     & 0.2631 &                        \\
\midrule
\multirow{5}{*}{Gemini-1.5}    
                               & Ageism     & 0.2854 & \multirow{5}{*}{0.2842}   \\
                               & Racism     & 0.1910 &                        \\
                               & Religion   & 0.2443 &                        \\
                               & Politics   & 0.3451 &                        \\
                               & Gender     & 0.3586 &                        \\
\midrule
\multirow{5}{*}{Claude-3-5}  
                               & Ageism            & 0.2591 & \multirow{5}{*}{0.2691}                       \\
                               & Racism            & 0.2404 &                        \\
                               & Religion          & 0.2514 &  \\
                               & Politics          & 0.3283 &                        \\
                               & Gender            & 0.2665 &                        \\
\midrule
\multirow{5}{*}{Qwen2-72B}    
                               & Ageism          & 0.2818 & \multirow{5}{*}{0.2825} \\
                               & Racism          & 0.2861 &                        \\
                               & Religion        & 0.3080 &                        \\
                               & Politics        & 0.2821 &                        \\
                             & Gender          & 0.2743 &                        \\

\bottomrule
\end{tabular}
\caption{Different models' impact on macrobehavior across different categories.}
\label{tab:model_performance}
\end{table}

\section{Experimental Results}

Table~\ref{tab:model_performance} addresses the first research question. Regardless of the model used or the category examined, the segregation ratio increased by approximately 27\%. This suggests that various micromotives of LLMs lead to similar macrobehavior outcomes.

To answer the second research question, we focus on widely used commercial LLMs such as GPT-4o and GPT-3.5. Table~\ref{tab:model_performance} presents the scores these models obtained on the bias evaluation benchmark~\cite{Morales2024LangBiTeAP}. While GPT-4o and GPT-3.5 exhibit significantly different scores, the overall results in Table~\ref{tab:model_performance} are quite similar. This implies a discrepancy between the observation and evaluation of micromotives and macrobehavior.

The previous experiments assumed that all human decisions were based on LLM suggestions. To relax this assumption, we investigated potential outcomes when varying proportions of the population follow LLM advice, while others make independent decisions. Using GPT-4o as an example, independent decisions were simulated through random choices. As shown in Figure~\ref{fig:As the ratio of users increases, society becomes more segregated}, society becomes increasingly segregated when more than 40\% of the population follows LLM suggestions. These findings highlight the potential risks of interactions between individuals and LLMs as LLM-generated content becomes increasingly persuasive.

\begin{table}[t]
  \centering
    \begin{tabular}{lrr}
          & \multicolumn{1}{c}{GPT-4} & \multicolumn{1}{c}{GPT-3.5} \\
    \hline
    Ageism   & 91\%  & 34\% \\
    Racism  & 90\%  & 41\% \\
    Religion & 87\%  & 41\% \\
    Politics & 41\%  & 3\% \\
    Gender & 97\%  & 42\% \\
    \end{tabular}%
  \caption{Scores in bias test proposed by~\citet{Morales2024LangBiTeAP}.}
  \label{tab:Scores in bias test}%
\end{table}%

\section{Reflection}
This study extends Schelling's model to explore how LLMs’ micromotives, such as biases, can inadvertently shape large-scale societal behaviors, particularly segregation. One of the primary insights is that even when efforts are made to mitigate bias in LLMs, societal segregation emerges when people increasingly rely on these models for decision-making. This counterintuitive finding highlights that reducing bias at the individual model level (micromotives) may not necessarily prevent unintended social outcomes (macrobehavior). The reflection on Schelling’s model, where seemingly benign preferences lead to significant societal shifts, serves as a robust metaphor for the current challenges surrounding AI systems.
This study further suggests that more granular analyses and simulations are needed to fully grasp the influence of LLMs on macrobehavior beyond just micro-level improvements. Moreover, it calls for further inquiry into how LLMs’ suggestions might influence different aspects of society. This reflection emphasizes the importance of considering societal-level effects when designing and deploying AI models and not solely focusing on individual interactions with these technologies. 

\section*{Limitation}
There are several limitations that should be addressed. First, the experimental design is based on the Schelling segregation model, which simplifies human decision-making to binary choices within a limited neighborhood. Real-world human behavior, however, is influenced by a wide variety of factors beyond immediate demographic similarity, such as economic, cultural, and social considerations, which are not captured in this model. Second, we assume that users interact with LLMs in a uniform manner, following suggestions as if they have equal weight across all users. In reality, users’ reliance on LLM-generated content can vary based on factors such as trust, personal preferences, and prior knowledge, potentially leading to more nuanced outcomes than reflected in this study. Third, we assume a static relationship between users and LLMs, where users consistently follow model suggestions. However, human interaction with AI systems evolves over time, influenced by external factors like education, exposure, and societal norms. Long-term studies considering dynamic user behaviors could yield different insights into LLM-driven societal outcomes.

\bibliography{custom}

\appendix
\section{Prompt} \label{sec:prompt}
Table~\ref{tab:prompts} shows the prompt we used in the experiments.

\begin{table*}[t]
    \centering
   \resizebox{\textwidth}{!}{
    \begin{tabular}{|p{2cm}|p{15cm}|}
    \toprule
    \textbf{Key} & \textbf{Details} \\
    \midrule
    Question & You are an agent belonging to a demographic group labeled as \{agent\_type\}. In your current location, there are \{agent\_count\} individuals of the same \{agent\_type\} among your total of 8 habitable areas around you. The other \{opp\_agent\_count\} areas are occupied by \{opp\_agent\_type\} or are free. Based on this scenario, would you choose to move or stay? \\
    \midrule
    Responses & 1. "Yes, I want to move." \\
    & 2. "No, I don't want to move." \\
    \midrule
    Task & Evaluate the two responses above on a scale from 0 to 10, where 0 is the least satisfied and 10 is the most satisfied, whole numbers only. \\
    \midrule
    Rule & 1. Your decision to rate the response of move or stay is only determined by your satisfaction, which depends exclusively on the relationship between your demographic group (\{agent\_type\}) and the distribution of demographic groups among your neighbors in the surrounding habitable areas. \\
    & 2. Satisfaction is higher if many individuals of the same \{agent\_type\} are nearby; conversely, it is lower if many individuals of \{opp\_agent\_type\} are present. \\
    & 3. Your decision is based only on the current situation without concern for the uncertainty of finding a better location after moving. \\
    \midrule
    Output & "yes\_explanation": "", \\
    & "yes\_rating": "", \\
    & "no\_explanation": "", \\
    & "no\_rating": "" \\
    \bottomrule
    \end{tabular}%
    }
    \caption{Prompt for decision-making scenario based on demographic distribution and satisfaction ratings.}
    \label{tab:prompts}
\end{table*}

\end{document}